\documentclass[aps,pra,twocolumn,groupedaddress,amsmath,amssymb]{revtex4}
\usepackage{graphicx}
\usepackage{dcolumn}
\begin{document}

\title{Generalized Measurement and Conclusive Teleportation with Nonmaximal Entanglement}
\author{Hyunjae Kim}
\author{Yong Wook Cheong}
\author{Hai-Woong Lee}
\affiliation{Department of Physics, Korea Advanced Institute of
Science and Technology, Daejeon 305-701, Korea}
\date{\today}
\begin{abstract}
We present linear optical schemes to perform generalized
measurements for conclusive teleportation when the sender and the
receiver share nonmaximal entanglement resulting from amplitude
errors during propagation or generation. Three different cases are
considered for which the states to be teleported are unknown
superpositions of (a) single-photon and vacuum states, (b)
vertically-polarized and horizontally-polarized photon states, and
(c) two coherent states of opposite phases. The generalized
measurement scheme for each case is analyzed, which indicates that
the  success probability is much more resistant to amplitude
errors for case (c) than for case (a) or (b).
\end{abstract}
\pacs{pacs} \maketitle

\newcommand{\bra}[1]{\left<#1\right|}
\newcommand{\ket}[1]{\left|#1\right>}
\newcommand{\abs}[1]{\left|#1\right|}

\section{Introduction}

Since its first proposal in 1993\cite{01}, quantum teleportation
has been the subject of intensive study, both theoretical and
experimental. Up to now, quantum teleportation of superposed
photon polarization states\cite{02}, superposed one-photon and
vacuum states\cite{03}, and continuous variable light
states\cite{04} has been demonstrated experimentally. Although
teleportation can be achieved ideally with the success probability
of 1 and the fidelity of 1, there exist many practical
limitations, such as various sources of decoherence, finite
efficiency of detectors, lack of reliable single-photon sources,
difficulty with perfect Bell-state measurements, and nonmaximal
entanglement, that need to be overcome before the experimental
performance can match ideal theoretical predictions.\\
\indent The process of teleportation may be regarded as the
transfer of quantum information through a quantum channel. When
the quantum channel is provided by a maximally entangled pair,
quantum teleportation faithfully transmits the quantum
information. Amplitude errors during generation or distribution of
entanglement, however, degrade the degree of entanglement. The
practical question of importance is thus what one should or can do
when the sender, Alice, and the receiver, Bob, share nonmaximal
entanglement. Different strategies should be adopted by Alice and
Bob depending upon the situation. They may choose the
teleportation scheme that would yield the maximum average
fidelity. In some cases, however, what is required is not to
obtain the maximum information but to avoid making any mistake
even at the expense of lowered success probability. In such cases,
they can adopt the strategy of ``conclusive
teleportation''\cite{05}, in which the teleportation succeeds with
the probability less than one but Alice knows when it succeeds or
not, and when it succeeds the fidelity of the teleported state is
one. There are three approaches to conclusive teleportation that
Alice and Bob can take. First, they may try to extract maximally
entangled pairs out of the given nonmaximally entangled pairs
through the process of entanglement concentration\cite{06}, and
perform teleportation with the concentrated pairs. The number of
the concentrated pairs is less than the number of the original
nonmaximally entangled pairs, i.e., the probability for
concentration is less than one, but the fidelity of the teleported
state using the concentrated pair as the quantum channel is one.
Second, Alice may perform the standard Bell-state measurements
directly with the nonmaximally entangled pairs and let Bob apply
an appropriate unitary transformation with the aid of an auxiliary
qubit to obtain the desired state probabilistically\cite{07}. The
success probability is less than one, but Bob knows whether he
obtained the desired state or not by observing the state of the
auxiliary qubit. Third, Alice may perform ``generalized
measurements''\cite{08}(or positive operator valued measurement)
upon the nonmaximally entangled pairs that distinguish
nonorthogonal ``Bell-type'' states conclusively with a certain
probability less than one\cite{05}. When the measurement outcome
corresponds to the conclusive event, the teleportation succeeds
with the fidelity of 1. \\
\indent It is this third approach we wish to study in detail here.
We construct linear optical schemes that enable one to perform the
desired generalized measurements for three different cases for
which two basis states for a qubit are given by (a)
$\left|1\right>$, a single-photon state, and $\left| 0 \right>$,
the vacuum state, (b) $\left| V \right>$, a vertically-polarized
photon state, and $\left|H \right>$, a horizontally-polarized
photon state, and (c) $\left| \alpha \right>$ and $\left| - \alpha
\right>$, two coherent light states with opposite phases. We
analyze and compare the generalized measurement process for
conclusive teleportation for the three cases. The reason we study
and compare these three cases is that, although the basic
principles governing the generalized measurement and teleportation
processes may be the same regardless of the choice of qubit
states, one may have significant advantages over another when the
actual experimental implementation of the scheme is considered. We
need only to recall that, in quantum key distribution, the choice
of $ \left| 1 \right\rangle$ and $\left| 0 \right\rangle$ has led
to the phase coding scheme which offers significant practical
advantages over the polarization coding scheme resulting from the
choice of $\left| V \right\rangle$ and $\left| H \right\rangle$.
In case of quantum teleportation with maximal entanglement,
investigations in the past have revealed that the three cases
require different levels of experimental sophistication in
Bell-state measurements and in state
transformations\cite{01,02,03,09,10,11,12}. For example,
polarizing beam splitters needed in Bell-state measurements for
case (b) are not required for case (a); instead, detectors in case
(a) need to distinguish a single photon from two\cite{09}. The
state transformations required in cases (a) and (b) are all
unitary, but some transformations involved in case (c) are
nonunitary\cite{11,12}. What makes case (c) still particularly
interesting is the fact that four ``quasi Bell states'' can in
principle be all distinguished with a high probability using only
linear optical means\cite{11,12}, whereas for cases (a) and (b)
only two of the four Bell states can be distinguished linear
optically. Since in general a different choice of qubit states
leads to a different experimental setup for generalized
measurements with different optical devices which may respond
differently to an attempt to compensate for amplitude errors that
caused nonmaximal entanglement, it should be of interest to
analyze and compare in detail the experimental schemes for
conclusive teleportation for different choices of qubit states.
Such comparison and analysis have indeed led to the main result of
our work that the success probability is much more resistant to
amplitude errors for case (c) than for case (a) or (b).

\section{Generalized Measurement and Conclusive Teleportation}

In this section we construct linear optical schemes for
generalized measurements and analyze the generalized measurements
that need to be performed with the proposed schemes, for the three
different cases (a), (b) and (c) mentioned in the previous
section.
\begin{figure}[b]
\includegraphics[width=4cm]{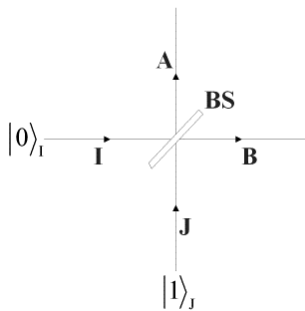}
\caption{\label{f1} Generation of the nonmaximally entangled state
$ \left(\cos\eta \left|1\right\rangle_A \left|0 \right\rangle_B
-\sin\eta \left| 0 \right\rangle_A \left| 1 \right\rangle_B
\right)$. {\bf BS} represents a beam splitter of transmission
coefficient $t= \cos \eta$ and reflection coefficient $r=\sin
\eta$, with input ports {\bf I} and {\bf J} and output ports {\bf
A} and {\bf B}. A single photon is incident on {\bf BS} through
the input port {\bf J}. }
\end{figure}

\subsection{ Conclusive teleportation of the state $\left( x \left|1
\right> + y \left| 0 \right>\right)$}

\begin{figure}
\includegraphics[width=7cm]{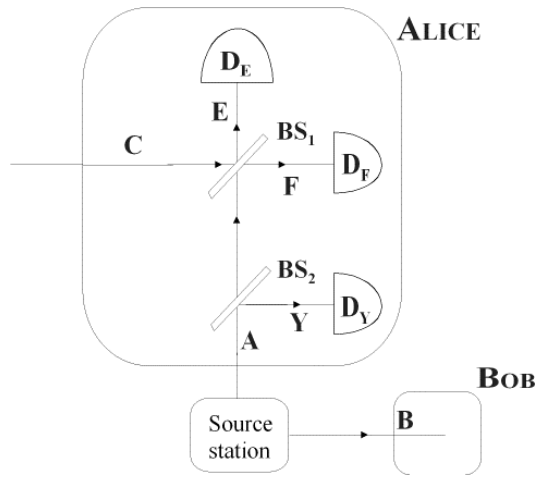}% Here is how to import EPS art
\caption{\label{f2} Linear optical scheme for a conclusive
teleportation of the state $\left( x \left| 1 \right\rangle_C + y
\left| 0 \right\rangle_C \right)$. The source station generates a
nonmaximally entangled state $\left| \Psi^- _\eta
\right\rangle_{AB} = \cos\eta \left| 1 \right\rangle_A \left| 0
\right\rangle_B - \sin \eta \left|0 \right\rangle_A \left| 1
\right\rangle_B$. $\bf D_E$, $\bf D_F$ and $\bf D_Y$ are
detectors, $\bf BS_1$ is a $50/50$ beam splitter, and $\bf BS_2$
is a beam splitter of transmission coefficient $t=\tan\eta$. }
\end{figure}
We first consider the case when two basis states for a qubit are
the single-photon and vacuum states, $\left| 1 \right>$ and
$\left| 0 \right>$. Let us suppose that Alice and Bob are given a
nonmaximally entangled state
\begin{equation}
\left| {\Psi _\eta ^ -  } \left(1,0\right) \right\rangle _{AB}
 \equiv \cos \eta \left| 1 \right\rangle _A \left| 0 \right\rangle _B  -
\sin \eta \left| 0 \right\rangle _A \left| 1 \right\rangle _B
,\label{e1}
\end{equation}
where the subscripts {\bf A} and {\bf B} signify that the beams
{\bf A} and {\bf B} that contain the entangled state belong to
Alice and Bob, respectively. The parameter $\eta$ determines the
degree of entanglement. Without loss of generality, we assume that
$\cos\eta
> \sin\eta$($ 0 < \eta < \frac{\pi}{4}$). The entangled state of
Eq.(\ref{e1}) results when a photon is passed through a beam
splitter whose transmission and reflection coefficients deviate
from $\frac{1}{\sqrt{2}}$. As shown in Fig.\ref{f1}, when a single
photon is incident on a beam splitter of transmission coefficient
$t=\cos\eta$ and reflection coefficient $r=\sqrt{1-t^2}=\sin\eta$,
the output state is given by Eq.(\ref{e1}).
\\
\indent Alice has another beam {\bf C} in an unknown state $
\left( x\left| 1 \right\rangle _C  + y\left| 0 \right\rangle _C
\right) \equiv { x \choose y} _C$
 with unknown
coefficients $x$ and $y$, which she wishes to teleport to Bob. The
total state of the beams {\bf A}, {\bf B} and {\bf C}, $ \left|
\Psi \right\rangle _{ABC}  = \left| {\Psi _\eta ^ -  }
\left(1,0\right) \right\rangle _{AB} { x \choose y} _C $, can be
expanded as (we adopt bracket notations for two-particle states
and vector notations for one-particle states)
\begin{equation}
\begin{array}{ll}
 \left| \Psi  \right\rangle _{ABC}  =&
 \frac{1}{2}\left[ {\left| {\Psi _\eta ^ -  } \right\rangle _{AC}
 \left( {\begin{array}{c}
   x  \\
   y  \\
\end{array}} \right)
 _B - \left| {\Psi _\eta ^ +  } \right\rangle _{AC}
 \left( {\begin{array}{*{20}c}
   x  \\
   { - y}  \\
\end{array}} \right)_B
} \right. \\
 & \left. {+\left| {\Phi _\eta ^ -  } \right\rangle _{AC}
 \left( {\begin{array}{c}
   y  \\
   x  \\
\end{array}} \right)_B }
 + \left| {\Phi _\eta ^ +  } \right\rangle _{AC} \left( {\begin{array}{c}
   -y  \\
   x  \\
\end{array}} \right)_B \right], \\ \label{e2}
 \end{array}
\end{equation}
where the four ``Bell-type states'' are defined by
\begin{eqnarray}
 \left| {\Psi _\eta ^ \mp  } \right\rangle _{AC} &\equiv&  \left| {\Psi _\eta ^ \mp \left(1,0 \right) } \right\rangle _{AC} \nonumber
 \\&=& \cos \eta \left| 1 \right\rangle _A \left| 0 \right\rangle _C  \mp \sin \eta \left| 0 \right\rangle _A \left| 1 \right\rangle _C ,
  \nonumber \\
 \left| {\Phi _\eta ^ \mp  } \right\rangle _{AC} &\equiv& \left| {\Phi _\eta ^ \mp  \left(1,0 \right) }
 \right\rangle _{AC} \nonumber \\
 &=& \cos \eta \left| 1 \right\rangle _A \left| 1 \right\rangle _C  \mp \sin \eta \left| 0 \right\rangle _A \left| 0
 \right\rangle _C .
 \label{e3}
 \end{eqnarray}
 When $\eta=\frac{\pi}{4}$, these four states reduce to the
 standard Bell states $\left| {\Psi  ^ {\mp}  } \right\rangle_{AC} =
 \frac{1}{\sqrt{2}} \left( \left| 1 \right\rangle_A \left| 0
 \right\rangle_C \mp \left| 0 \right\rangle_A \left| 1
 \right\rangle_C \right)$ and $\left| {\Phi  ^ {\mp}  } \right\rangle_{AC} =
 \frac{1}{\sqrt{2}}\left( \left| 1 \right\rangle_A \left| 1
 \right\rangle_C \mp \left| 0 \right\rangle_A \left| 0
 \right\rangle_C \right)$. \\
 \indent The conclusive teleportation protocol
 calls for a generalized measurement which unambiguously
 distinguishes at least two of the four nonorthogonal states of
 Eq.(\ref{e3}) at the price of occasional failures\cite{05}. We
 recall that the linear optical scheme that distinguishes two of
 the standard Bell states $\left| \Psi ^\mp \right\rangle$ and $\left| \Phi
 ^\mp \right\rangle$ consists of a $50/50$ beam splitter and two
 detectors ( $\bf BS_1$, $\bf D_E$ and $\bf D_F$ of Fig.\ref{f2})\cite{09}. The
 generalized measurement to distinguish unambiguously two of the
 four states of Eq.(\ref{e3}) can be accomplished by adding to
 them another beam splitter of transmission coefficient $t=\tan
 \eta$ and a third detector ($\bf BS_2$ and $\bf D_Y$ of Fig.\ref{f2}) in the
 path of the beam {\bf A}, as shown in Fig.\ref{f2}. By utilizing the
 relations between photon creation operators of different modes
 given by $ A^\dag = \frac{1}{\sqrt{2}} \left( t E^\dag -t F^\dag
 \right) -r Y^\dag$, $C^\dag = \frac{1}{\sqrt{2}}\left( E^\dag +F^\dag
 \right)$, where $r$ is the reflection coefficient $r= \sqrt{1-\tan^2
 \eta}$, we obtain that the state $\left| \Psi \right\rangle_{ABC}
 = \left| \Psi^- _\eta \right\rangle_{AB} {x \choose y}_C$ is transformed, via the action of the beam splitters
 $\bf BS_1$ and $\bf BS_2$, into $\left| \Psi \right\rangle_{EFYB}$,
 where
\begin{widetext}
 \begin{equation}
\begin{array}{ll}
 \left| \Psi  \right\rangle _{EFYB}  =&  \frac{1}{2}\left\{ {\left( { - \sqrt 2 \sin \eta \left| 0
\right\rangle _E \left| 1 \right\rangle _F \left| 0 \right\rangle
_Y  - \sqrt {\cos 2\eta } \left| 0 \right\rangle _E \left| 0
\right\rangle _F \left| 1 \right\rangle _Y } \right)\left(
{\begin{array}{c}
   x  \\
   y  \\
\end{array}} \right)_B } \right. \\
&  - \left( {\sqrt 2 \sin \eta \left| 1 \right\rangle _E \left| 0
\right\rangle _F \left| 0 \right\rangle _Y  - \sqrt {\cos 2\eta }
\left| 0 \right\rangle _E \left| 0 \right\rangle _F \left| 1
\right\rangle _Y } \right)\left( {\begin{array}{c}
   x  \\
   { - y}  \\
\end{array}} \right)_B  \\
&  + \left( {\sin \eta \left| {\Psi ^ -  \left( {2,0} \right)}
\right\rangle _{EF} \left| 0 \right\rangle _Y  - \sqrt {\cos 2\eta
} \left| {\Psi ^ +  \left( {1,0} \right)} \right\rangle _{EF}
\left| 1 \right\rangle _Y  - \sin \eta \left| 0 \right\rangle _E
\left| 0 \right\rangle _F \left| 0 \right\rangle _Y }
\right)\left( {\begin{array}{c}
   y  \\
   x  \\
\end{array}} \right)_B  \\
& \left. { + \left( {\sin \eta \left| {\Psi ^ -  \left( {2,0}
\right)} \right\rangle _{EF} \left| 0 \right\rangle _Y  - \sqrt
{\cos 2\eta } \left| {\Psi ^ +  \left( {1,0} \right)}
\right\rangle _{EF} \left| 1 \right\rangle _Y  + \sin \eta \left|
0 \right\rangle _E \left| 0 \right\rangle _F \left| 0
\right\rangle _Y } \right)\left( {\begin{array}{c}
   -y  \\
   x  \\
\end{array}} \right)_B } \right\} ,\\
 \end{array}
\label{e4}
 \end{equation}
\end{widetext}
where the state $\left| \Psi^- \left(2,0\right)
\right\rangle_{EF}$ denotes the state $\frac{1}{\sqrt{2}} \left(
\left| 2 \right\rangle_E \left| 0 \right\rangle_F - \left|0
\right\rangle_E \left| 2 \right\rangle_F \right)$.

 The generalized
measurement consists of measuring the number of photons that
arrive at each of the three detectors $\bf D_E$, $\bf D_F$ and
$\bf D_Y$. It is clear from Eq.(\ref{e4}) that the teleportation
succeeds either if $\bf D_F$ registers one photon and $\bf D_E$
and $\bf D_Y$ none or if $\bf D_E$ registers one photon and $\bf
D_F$ and $\bf D_Y$ none. All other measurement results correspond
to inconclusive events. It is also clear from Eq.(\ref{e4}) that
the success probability for the conclusive teleportation is
$P=\sin^2 \eta$. When $\eta=\frac{\pi}{4}$, i.e., when Alice and
Bob share maximal entanglement, $P=\frac{1}{2}$ consistent with
the previous analysis\cite{09}. As the degree of entanglement
decreases, the success probability decreases, too. Nevertheless,
the fidelity $F$ remains 1. We note that a successful conclusive
teleportation for the present case requires detectors $\bf D_E$
and $\bf D_F$ to discriminate between 0, 1 and 2 photons and
detector $\bf D_Y$ to discriminate between 0 and 1 photon.

\subsection{Conclusive teleportation of the state $\left( x \left|V
\right\rangle +y \left| H \right\rangle \right)$}

\begin{figure}
\includegraphics[width=7cm]{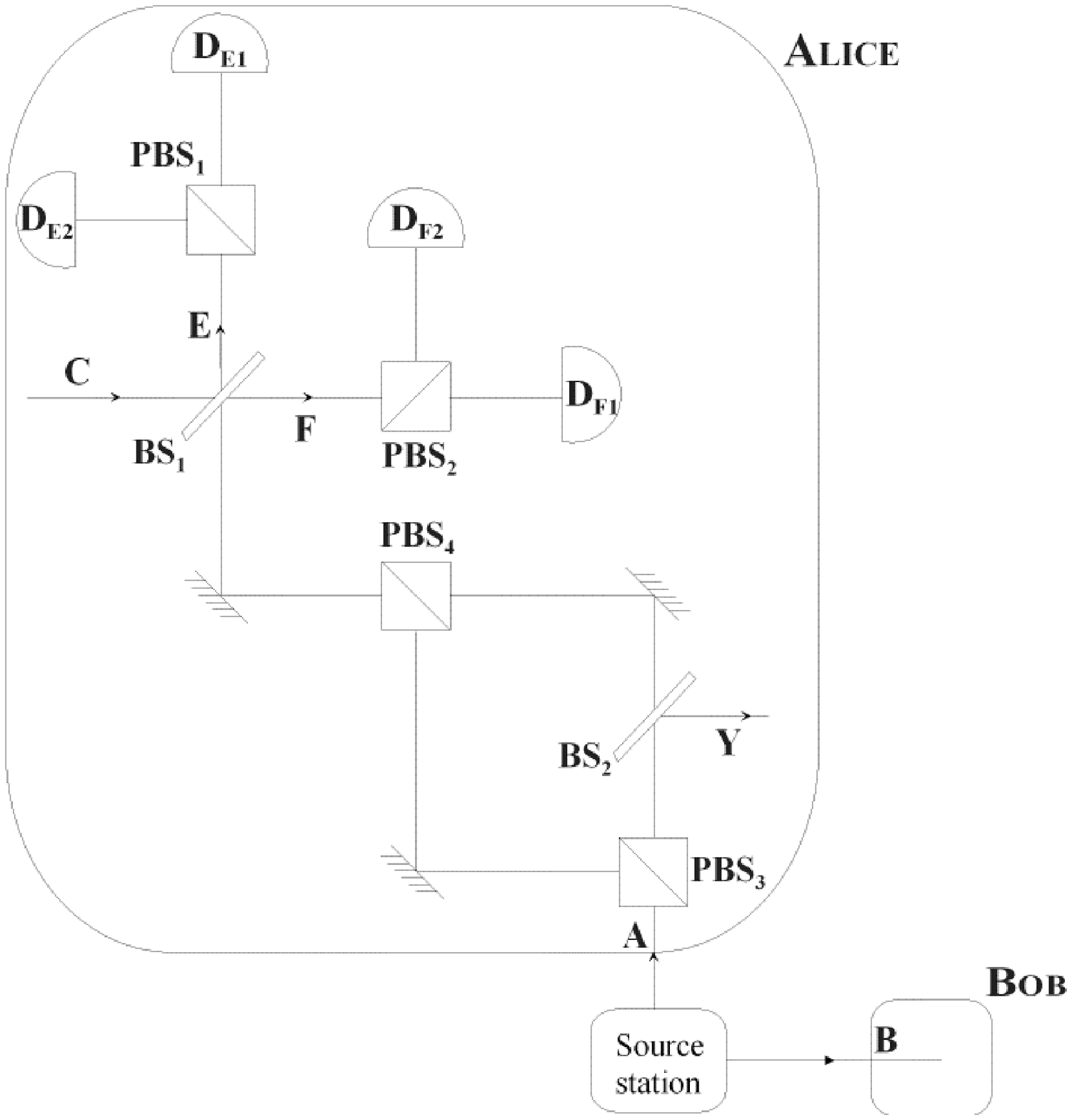}% Here is how to import EPS art
\caption{\label{f3} Linear optical scheme for a conclusive
teleportation of the state $\left( x \left| V \right\rangle_C +y
\left|H \right\rangle_C \right)$. The source station generates a
nonmaximally entangled state $ \left| \Psi^- _\eta
\right\rangle_{AB} =\cos \eta \left|V \right\rangle_A \left| H
\right\rangle_B -\sin \eta \left|H \right\rangle_A \left| V
\right\rangle_B$. $\bf D_{E1}$, $\bf D_{E2}$, $\bf D_{F1}$, and
$\bf D_{F2}$ are detectors, $\bf PBS_1$, $\bf PBS_2$, $\bf PBS_3$
and $\bf PBS_4$ are polarizing beam splitters, $\bf BS_1$ is a
$50/50$ beam splitter, and $\bf BS_2$ is a beam splitter of
transmission coefficient $t=\tan \eta$. }
\end{figure}

We next consider the case when two basis states for a qubit are
the vertically-polarized and horizontally-polarized photon states
$ \left| V \right\rangle$ and $ \left| H \right\rangle$. The
nonmaximally entangled state $\left| \Psi_\eta ^- \left(V,H
\right) \right\rangle_{AB} $ for two photons {\bf A} and {\bf B}
now takes the form of Eq.(\ref{e1}) with $\left| 1 \right\rangle$
and $\left| 0 \right\rangle$ replaced by $\left| V \right\rangle$
and $\left| H \right\rangle$. This type of nonmaximally entangled
state can be produced by using a spontaneous parametric
down-conversion source which consists of two crystals with their
optic axis oriented orthogonal to each other, as demonstrated by
White {\it et al}\cite{13}. The unknown polarization state of the
photon that Alice wishes to teleport to Bob is $ \left( x \left| V
\right\rangle_C + y \left| H \right\rangle_C \right)$, which we
again denote by ${x \choose y}_C $. The four Bell-type states that
need to be distinguished are now
\begin{eqnarray}
 \left| {\Psi _\eta ^ \mp  } \right\rangle _{AC} &\equiv& \left| {\Psi _\eta ^ \mp  } \left(V,H \right) \right\rangle _{AC} \nonumber\\
   &=& \cos \eta \left| V \right\rangle _A \left| H \right\rangle _C  \mp \sin \eta \left| H \right\rangle _A \left| V \right\rangle _C ,
  \nonumber \\
 \left| {\Phi _\eta ^ \mp  } \right\rangle _{AC} &\equiv& \left| {\Phi _\eta ^ \mp  } \left(V,H \right) \right\rangle _{AC} \nonumber\\
 &=& \cos \eta \left| V \right\rangle _A \left| V \right\rangle _C  \mp \sin \eta \left| H \right\rangle _A \left| H \right\rangle _C .
 \label{e5}
\end{eqnarray}
\indent We recall that the linear optical scheme that
distinguishes two of the standard Bell states  $\left| {\Psi
^{\mp}  } \right\rangle = \frac{1}{\sqrt{2}} \left( \left| V
\right\rangle
 \left| H \right\rangle   \mp  \left| H \right\rangle
 \left| V \right\rangle \right)$ and $ \left| {\Phi  ^{\mp}
} \right\rangle   = \frac{1}{\sqrt{2}} \left(\left| V
\right\rangle \left| V \right\rangle \right. \left.  \mp  \left| H
\right\rangle \left| H \right\rangle \right)$ consists of a
$50/50$ beam splitter, two polarizing beam splitters and four
detectors ( $\bf BS_1$, $\bf PBS_1$, $\bf PBS_2$, $\bf D_{E1}$,
$\bf D_{E2}$, $\bf D_{F1}$ and $\bf D_{F2}$ of
Fig.\ref{f3})\cite{10}. The corresponding generalized measurement
that distinguishes two of the states of Eq.(\ref{e5}) can be
accomplished by adding another beam splitter of transmission
coefficient $t=\tan\eta$ and two additional polarizing beam
splitters ($\bf BS_2$, $\bf PBS_3$ and $\bf PBS_4$ of
Fig.\ref{f3}) in the path of photon {\bf A}, as shown in
Fig.\ref{f3}. The polarizing beam splitters transmit
vertically-polarized photons and reflect horizontally-polarized
photons. The combination of $\bf PBS_3$ and $\bf BS_2$ thus acts
to reduce the amplitude of the vertically-polarized component of
wave {\bf A} by a factor of $t=\tan\eta$, while leaving the
amplitude of the horizontally-polarized component unaltered. In
fact, one obtains through a straightforward calculation that the
state $ \left| \Psi \right\rangle _{ABC}  = \left| \Psi_\eta ^-
\left( V,H \right) \right\rangle_{AB} {x \choose y }_C$ is
transformed, via the action of $\bf PBS_3$, $\bf BS_2$, $\bf
PBS_4$ and $\bf BS_1$, into the state $\left| \Psi
\right\rangle_{EFYB}$, where
\begin{widetext}
\begin{equation}
\begin{array}{ll}
 \left| \Psi  \right\rangle _{EFYB}  =&
 \frac{1}{2}\left\{ {\left[ {\sqrt 2 \sin \eta \left| {\Psi ^ -  \left( {V,H} \right)} \right\rangle _{EF} \left| 0 \right\rangle _Y  +
 \sqrt {\cos 2\eta } \left| {\Psi ^ +  \left( {H,0} \right)} \right\rangle _{EF} \left| V \right\rangle _Y } \right]\left( {\begin{array}{*{20}c}
   x  \\
   y  \\
\end{array}} \right)_B } \right. \\
&  - \left[ {\sqrt 2 \sin \eta \left| {\Psi ^ -  \left( {VH,0}
\right)} \right\rangle _{EF} \left| 0 \right\rangle _Y  + \sqrt
{\cos 2\eta } \left| {\Psi ^ +  \left( {H,0} \right)}
\right\rangle _{EF} \left| V \right\rangle _Y } \right]\left(
{\begin{array}{*{20}c}
   x  \\
   { - y}  \\
\end{array}} \right)_B  \\
&  + \left[ {\sin \eta \left( {\left| {\Psi ^ -  \left( {2V,0}
\right)} \right\rangle _{EF}  - \left| {\Psi ^ -  \left( {2H,0}
\right)} \right\rangle _{EF} } \right)\left| 0 \right\rangle _Y  +
\sqrt {\cos 2 \eta } \left| {\Psi ^ +  \left( {V,0} \right)}
\right\rangle _{EF} \left| V \right\rangle _Y } \right]\left(
{\begin{array}{*{20}c}
   y  \\
   x  \\
\end{array}} \right)_B  \\
& \left. { + \left[ {\sin \eta \left( {\left| {\Psi ^ -  \left(
{2V,0} \right)} \right\rangle _{EF}  + \left| {\Psi ^ -  \left(
{2H,0} \right)} \right\rangle _{EF} } \right)\left| 0
\right\rangle _Y  + \sqrt {\cos 2 \eta } \left| {\Psi ^ +  \left(
{V,0} \right)} \right\rangle _{EF} \left| V \right\rangle _Y }
\right]\left( {\begin{array}{*{20}c}
   { - y}  \\
   x  \\
\end{array}} \right)_B } \right\} ,\label{e6}
 \end{array}
\end{equation}
\end{widetext}
The Bell-state notations used in Eq.(\ref{e6}) should be
self-evident. For example, $\left| \Psi^- \left(VH,0 \right)
\right\rangle_{EF}= \frac{1}{\sqrt{2}} \left( \left| VH
\right\rangle_E \left| 0 \right\rangle_F - \left| 0
\right\rangle_E \left| VH \right\rangle_F \right)$, where $\left|
0 \right\rangle_E$ means that there is neither
vertically-polarized nor horizontally-polarized photon of mode
{\bf E}, and $\left| VH \right\rangle_E$ stands for the state of
one vertically-polarized photon and one horizontally-polarized
photon of mode {\bf E}. Similarly,  $\left| \Psi^- \left(2V,0
\right) \right\rangle_{EF}= \frac{1}{\sqrt{2}} \left( \left| 2V
\right\rangle_E \left| 0 \right\rangle_F - \left| 0
\right\rangle_E \left| 2 V \right\rangle_F \right)$, where $\left|
2V \right\rangle_E$ stands for the state of two
vertically-polarized photons of mode {\bf E}. It is clear from
Eq.(\ref{e6}) that the teleportation succeeds if detectors $\bf
D_{E1}$ and $\bf D_{F2}$ register a photon each or if detectors
$\bf D_{E2}$ and $\bf D_{F1}$ register a photon each
(corresponding to the term $\left| \Psi^- \left(V,H \right)
\right\rangle _{EF}$ in the square bracket of the first line of
Eq.(\ref{e6})), or if detectors $\bf D_{E1}$ and $\bf D_{E2}$
register a photon each, or if detectors $\bf D_{F1}$ and $\bf
D_{F2}$ register a photon each (corresponding to the term $\left|
\Psi^- \left(VH,0 \right) \right\rangle _{EF}$ in the square
bracket of the second line of Eq.(\ref{e6})). All other
measurement results correspond to inconclusive events. It is also
clear from Eq.(\ref{e6}) that, as in the previous case where the
basis states for a qubit are the single-photon and vacuum states,
the success probability is $\sin^2 \eta$ and the fidelity is 1.
The present case, however, holds an interesting advantage over the
previous case in that there is no need for a detector to detect
photons of mode $\bf Y$. The combination $\left| V \right\rangle_E
\left| H \right\rangle_F$, $\left| H \right\rangle_E \left| V
\right\rangle_F$, $\left| V \right\rangle_E \left| H
\right\rangle_E$, or $\left| V \right\rangle_F \left| H
\right\rangle_F$ does not appear in the third or fourth line of
Eq.(\ref{e6}). Another important advantage is that, in the present
case, detectors do not need to distinguish between one and two
photons. For example, a click each at detectors $\bf D_{E1}$ and
$\bf D_{F2}$ along with no detection of photons at detectors $\bf
D_{E2}$ and $\bf D_{F1}$ indicates conclusively that the state of
photon {\bf B} is ${x \choose y}_B =  x \left| V \right\rangle_B +
y \left| H \right\rangle_B $.

\subsection{Conclusive teleportation of the state $\left( x \left| \alpha
\right\rangle + y \left| - \alpha \right\rangle \right)$}

\begin{figure}
\includegraphics[width=4cm]{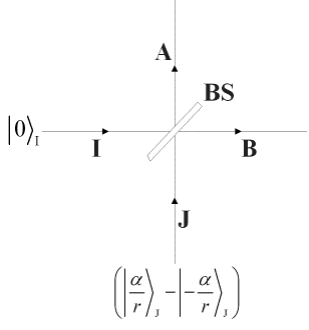}% Here is how to import EPS art
\caption{\label{f4} Generation of the nonmaximally entangled state
$N^- _{\alpha \beta} \left( \left| \alpha \tan\eta \right\rangle_A
\left|- \alpha \right\rangle_B -\left| - \alpha \tan\eta
\right\rangle_A \left|\alpha\right\rangle_B \right)$. {\bf BS}
represents a beam splitter of transmission coefficient $t=\sin
\eta$ and reflection coefficient $r=\cos \eta$, with input ports
{\bf I} and {\bf J} and output ports {\bf A} and {\bf B}. The
input port {\bf J} is illuminated with light in the state $\left(
\left| {\alpha \over {\cos \eta}} \right\rangle_J - \left| -{
\alpha \over{\cos \eta}} \right\rangle_J \right)$. }
\end{figure}

Finally, we consider the case when two basis states for a qubit
are the coherent states of opposite phases, $\left| \alpha
\right\rangle$ and $\left| - \alpha \right\rangle$. The two basis
states in this case are not orthogonal, $ \left\langle {\alpha }
 \mathrel{\left | {\vphantom {\alpha  { - \alpha }}}
 \right. \kern-\nulldelimiterspace}
 {{ - \alpha }} \right\rangle  = e^{ - 2\left| \alpha  \right|^2 }
 $; but, if $ \left| \alpha \right|$ is not too small ($
\left| \alpha  \right| \gtrsim 3 $), the overlap is sufficiently
small that they can be treated orthogonal without much error. It
is important to note that an amplitude loss experienced by light
in a coherent state $\left| \alpha \right\rangle$ reduces not the
probability amplitude associated with the state but the parameter
$\alpha$ itself\cite{14}. The nonmaximally entangled state we
consider here can thus be written as
\begin{equation}
\left| \Psi^- _\eta \right\rangle_{AB} = N^- _{\alpha \beta}
\left( \left| \beta \right\rangle_A \left| - \alpha
\right\rangle_B - \left| - \beta \right\rangle_A \left| \alpha
\right\rangle_B \right) , \label{e7}
\end{equation}
where $N^- _{\alpha \beta} $ is the normalization constant $
\left( N^- _{\alpha \beta} =   1/ \sqrt{2 \left[ 1- e^{ -2 \left(
 \left| \alpha \right|^2 + \left| \beta \right|^2 \right) } \right] } \right) $. We assume without loss of generality that
$\alpha$ and $\beta$ are real and that $\frac{\beta}{\alpha}
\equiv \tan\eta < 1$. The nonmaximally entangled state of
Eq.(\ref{e7}) can be generated by illuminating a beam splitter of
transmission coefficient $t= \sin\eta$ and reflection coefficient
$r= \cos \eta$ through the input port {\bf J} with light in the
state $\left( \left| \frac{\alpha}{\cos\eta} \right\rangle_J -
\left| -{ \alpha \over {\cos\eta} }\right\rangle_J \right)$, as
shown in Fig.\ref{f4}.

The unknown state of light wave {\bf C} that Alice wants to
teleport to Bob is $ N_{\alpha x y } \left( x \left| \alpha
\right\rangle_C +y \left| - \alpha \right\rangle_C \right) \equiv
N_{\alpha x y } { x \choose y}_C$, where $N_{\alpha x y}$ is the
normalization constant $ \left( N_{\alpha x y}= 1/ \sqrt{ \left| x
\right|^2 + \left| y \right|^2 + \left(x^* y + x y^* \right) e^{-2
\left| \alpha \right|^2 }}\right)$. The total state of the light
waves {\bf A}, {\bf B} and {\bf C}, $ \left| \Psi
\right\rangle_{ABC}
 = \left| \Psi^- _\eta \right\rangle_{AB} N_{\alpha x y} {x \choose y}_C$, can be expanded as
\begin{equation}
\begin{array}{ll}
 \left| \Psi  \right\rangle _{ABC} = &\frac{{N_{\alpha \beta }^ -  N_{\alpha xy} }}{2}  \times\\
& \left\{ \left( {\left| \beta  \right\rangle _A \left| { - \alpha
} \right\rangle _C  - \left| { - \beta } \right\rangle _A \left|
\alpha  \right\rangle _C } \right)
\left(\begin{array}{c} x\\y
\end{array}\right)_B
 \right. \\
 & - \left( {\left| \beta  \right\rangle _A \left| { - \alpha } \right\rangle _C  +
 \left| { - \beta } \right\rangle _A \left| \alpha  \right\rangle _C } \right)
 \left(\begin{array}{c}
x\\-y
\end{array}\right)_B \\
 & + \left( {\left| \beta  \right\rangle _A \left| \alpha  \right\rangle _C  -
 \left| { - \beta } \right\rangle _A \left| { - \alpha } \right\rangle _C } \right)
 \left(\begin{array}{c}
y\\x
\end{array}\right)_B \\
 &\left.  + \left( {\left| \beta  \right\rangle _A \left| \alpha  \right\rangle _C
 + \left| { - \beta } \right\rangle _A \left| { - \alpha } \right\rangle _C } \right)
 \left(\begin{array}{c}
 -y\\x
 \end{array}\right)_B
\right\} \\
 \end{array}
\label{e8}
\end{equation}
The four Bell-type states which need to be distinguished here are
$\left( \left| \beta \right\rangle_A \left| - \alpha \right\rangle
_C \mp \left| - \beta \right\rangle_A \left| \alpha
\right\rangle_C \right)$ and $\left( \left| \beta \right\rangle_A
\left|  \alpha \right\rangle _C \mp \left| - \beta \right\rangle_A
\left| -\alpha \right\rangle_C \right)$.

\begin{figure}[t]
\includegraphics[width=7cm]{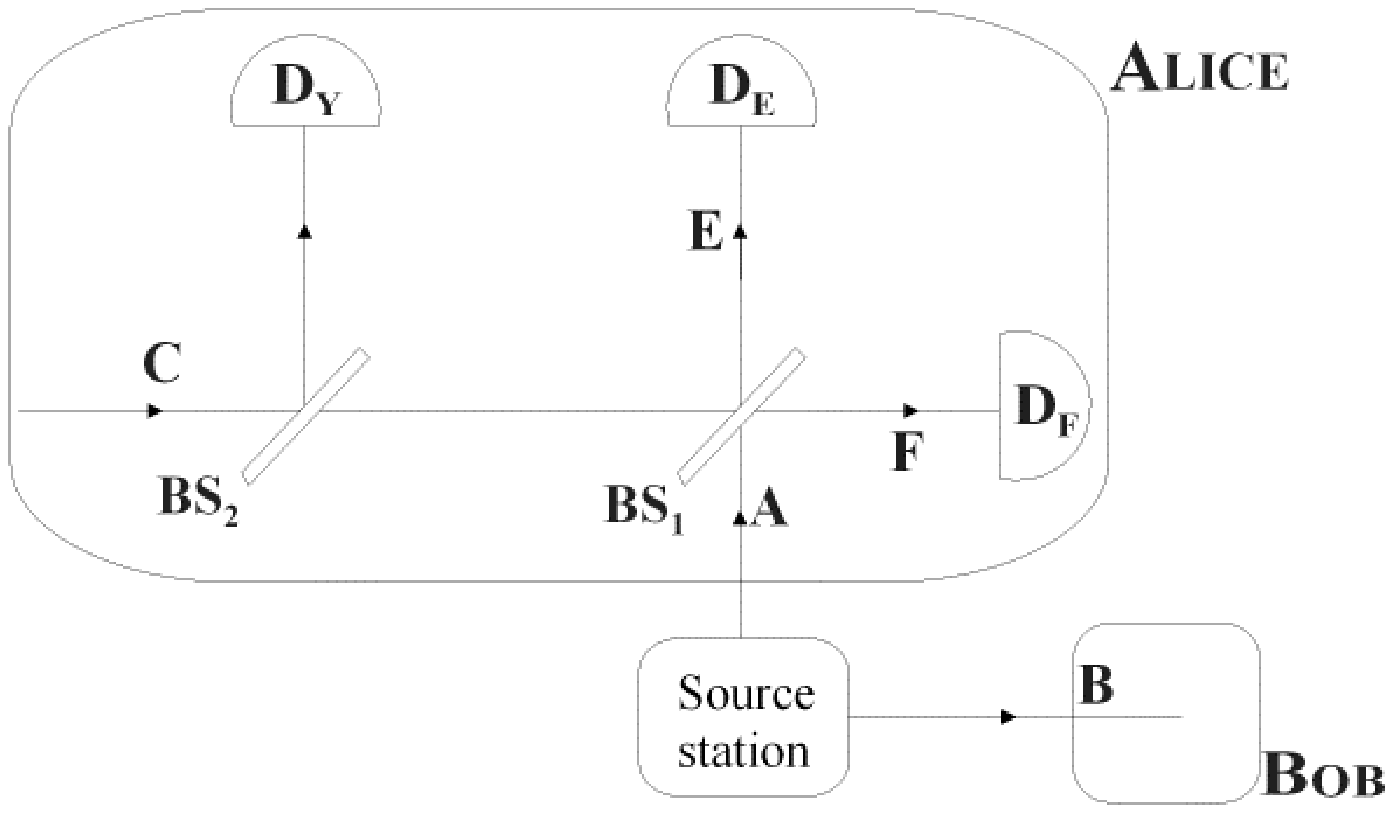}
\caption{\label{f5} Linear optical scheme for a conclusive
teleportation of the state $\left( x \left| \alpha \right\rangle_C
+ y \left| - \alpha \right\rangle_C\right)$. The source station
generates a nonmaximally entangled state $\left| \Psi^- _\eta
\right\rangle_{AB} = N^- _{\alpha \beta} \left( \left| \beta
\right\rangle_A \left| - \alpha \right\rangle_B - \left|-\beta
\right\rangle_A \left| \alpha \right\rangle_B \right)$, where
$\beta= \alpha \tan \eta$. $\bf D_E$, $\bf D_F$, and $\bf D_Y$ are
detectors, $\bf BS_1$ is a $50/50$ beam splitter, and $\bf BS_2$
is a beam splitter of transmission coefficient $t=\tan \eta$. }
\end{figure}

We recall that the four ``quasi-Bell states'' $\left( \left|
\alpha \right\rangle_A \left| - \alpha \right\rangle _C \mp \left|
- \alpha \right\rangle_A \left| \alpha \right\rangle_C \right)$
and $\left( \left| \alpha \right\rangle_A \left|  \alpha
\right\rangle _C \right. \mp \left. \left| - \alpha
\right\rangle_A \left| -\alpha \right\rangle_C \right)$ can all be
distinguished (with a high probability if $\alpha$ is not too
small) by a linear optical scheme using a $50/50$ beam splitter
and two detectors ($\bf BS_1$, $\bf D_E$ and $\bf D_F$ of
Fig.\ref{f5})\cite{11,12}, provided that the two detectors $\bf
D_E$ and $\bf D_F$ can discriminate between odd and even numbers
of photons. The corresponding generalized measurement to
distinguish all four quasi-Bell-type states $\left( \left| \beta
\right\rangle \left| - \alpha \right\rangle \mp \left| - \beta
\right\rangle \left| \alpha \right\rangle \right)$ and $\left(
\left| \beta \right\rangle \left| \alpha \right\rangle \mp \left|
- \beta \right\rangle \left| -\alpha \right\rangle \right)$ can be
accomplished by adding a second beam splitter of transmission
coefficient $t =\tan\eta$ and an additional detector that can also
discriminate between odd and even photons ($\bf BS_2$ and $\bf
D_Y$ of Fig.\ref{f5}) in the path of light wave C, as shown in
Fig.\ref{f5}. A straightforward calculation yields that the state
$\left| \Psi \right\rangle_{ABC}$ of Eq.(\ref{e8}) is transformed,
via the action of $\bf BS_1$ and $\bf BS_2$, into $\left| \Psi
\right\rangle_{EFYB}$, where
\begin{widetext}
\begin{equation}
\begin{array}{ll}
 \left| \Psi  \right\rangle _{EFYB}  = \frac{{N_{\alpha \beta }^ -  N_{\alpha xy} }}{2}
&
  \left\{ {\left| 0 \right\rangle _E \left( {\left| { - \sqrt 2 \beta } \right\rangle _F \left| { - r\alpha } \right\rangle _Y  - \left| {\sqrt 2 \beta } \right\rangle _F \left| {r\alpha } \right\rangle _Y } \right)\left(\begin{array}{c} x \\y \end{array}\right)_B} \right. \\
 & - \left| 0 \right\rangle _E \left( {\left| { - \sqrt 2 \beta } \right\rangle _F \left| { - r\alpha } \right\rangle _Y  + \left| {\sqrt 2 \beta } \right\rangle _F \left| {r\alpha } \right\rangle _Y } \right) \left(\begin{array}{c} x \\-y \end{array}\right)_B\\
  &+ \left| 0 \right\rangle _F \left( {\left| {\sqrt 2 \beta } \right\rangle _E \left| {r\alpha } \right\rangle _Y  - \left| { - \sqrt 2 \beta } \right\rangle _E \left| { - r\alpha } \right\rangle _Y } \right) \left(\begin{array}{c} y \\x \end{array}\right)_B\\
 & \left. { + \left| 0 \right\rangle _F \left( {\left| {\sqrt 2 \beta } \right\rangle _E \left| {r\alpha } \right\rangle _Y  + \left| { - \sqrt 2 \beta } \right\rangle _E \left| { - r\alpha } \right\rangle _Y } \right) \left(\begin{array}{c} -y \\x \end{array}\right)_B} \right\} \\
 \end{array}
\label{e9}
\end{equation}
\end{widetext}
where $r= \sqrt{1-\tan^2 \eta}$. Noting that $ \left( \left| u
\right\rangle_F \left| v \right\rangle_Y  \mp  \left| -u
\right\rangle_F \left| -v \right\rangle_Y \right) $ can be written
as $\frac{1}{2} \left[ \left( \left| u \right\rangle_F \mp \left|
-u \right\rangle_F \right) \left( \left| v \right\rangle_Y +
\left| -v \right\rangle_Y\right) \right. + \left. \left( \left| u
\right\rangle_F \pm \left| -u \right\rangle_F\right)\left( \left|
v \right\rangle_Y - \left| -v \right\rangle_Y\right)\right] $, all
four cases represented by the four lines of Eq.(\ref{e9}) can be
discriminated by observing which of the three detectors $\bf D_E$,
$\bf D_F$ and $\bf D_Y$ registers zero, odd or even number of
photons. The possible measurement outcomes are summarized in Table
\ref{t1}.
\begin{table}[t]
\caption{Possible outcomes of the detectors $\bf D_E$, $\bf D_F$
and $\bf D_Y$ and corresponding states of light wave {\bf
B}}\label{t1}
\begin{ruledtabular}
\begin{tabular}{cccc}
 \hline
 $\bf D_E$ & $\bf D_F$ & $\bf D_Y$ & State of {\bf B} \\ \hline\hline
  0 & odd & even & $x \left| \alpha \right\rangle_B + y \left| -
 \alpha \right\rangle_B$ \\
  0 & even & odd &  \\ \hline
  0 & even & even & $x \left| \alpha \right\rangle_B - y \left| -
 \alpha \right\rangle_B$ \\
  0 & odd & odd & \\ \hline
  odd & 0 & even & $x \left| -\alpha \right\rangle_B + y \left|
 \alpha \right\rangle_B$ \\
  even & 0 & odd &  \\ \hline
  even & 0 & even & $x \left| -\alpha \right\rangle_B + y \left|
 \alpha \right\rangle_B$ \\
  odd & 0 & odd &  \\ \hline
\end{tabular}
\end{ruledtabular}
\end{table}
The only case in which the generalized measurement fails
is when the two detectors $\bf D_E$ and $\bf D_F$ both register
zero photon. The success probability $P$ is thus given by
($P=1-$Probability for both $\bf D_E$ and $\bf D_F$ to detect no
photon). While $P=\sin^2 \eta$ for the previous two cases (a) and
(b), the success probability for the present case will show a
different behavior with respect to $\eta$. Details of the
comparison will be given in the next section.

\section{Discussion}

\begin{figure}[t]
\includegraphics[width=7cm]{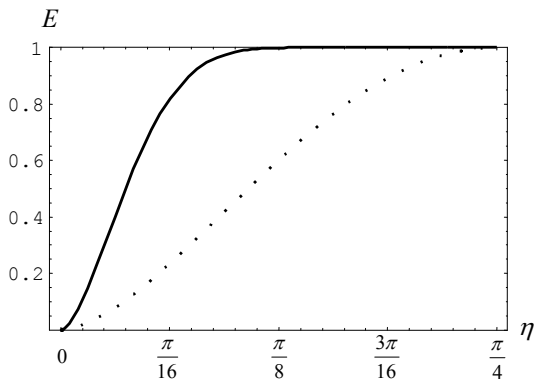}% Here is how to import EPS art
\caption{\label{f7} The degree of entanglement $E$ vs.~$\eta$ for
state (\ref{e7}) (solid curve) and for state (\ref{e1}) (dotted
curve). }
\end{figure}

\begin{figure}
\includegraphics[width=7cm]{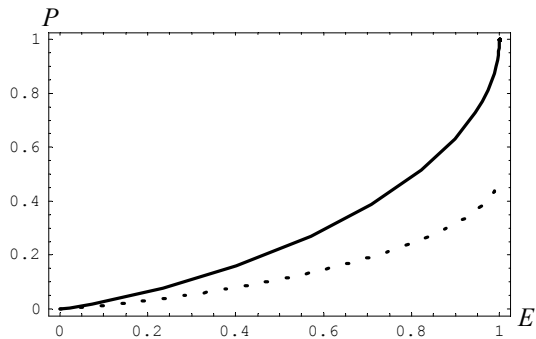}% Here is how to import EPS art
\caption{\label{f8} The success probability $P$ vs.~the degree of
entanglement $E$ for case (c) (solid curve) and for cases (a) and
(b) (dotted curve) }
\end{figure}

We have presented linear optical schemes for conclusive
teleportation when Alice and Bob share nonmaximal entanglement.
Considered in detail are three different cases when two basis
states for a qubit are (a) $\left| 1 \right\rangle$ and $\left| 0
\right\rangle$, (b) $\left|V\right\rangle$ and $\left| H
\right\rangle$, and (c) $\left| \alpha \right\rangle$ and $\left|
- \alpha \right\rangle$. The essence of the schemes common to all
three cases is the use of a beam splitter of an appropriate value
of transmission coefficient to reduce the amplitude of one part of
the entangled state, consistent with the idea of loss-induced
generalized measurement\cite{08}. There exist, however, some
interesting differences in details of the schemes between the
three cases. We have found that, while for case (a) an additional
detector that can distinguish between zero and one photon is
needed in addition to the beam splitter, for case (b) no
additional detector is required. The success probability in both
cases (a) and (b) decreases as $\sin^2 \eta$ as  the degree of
entanglement is decreased. In contrast to this, the success
probability for case (c) behaves in a more complicated way, as the
generalized measurement in this case fails only when the two
detectors $\bf D_E$ and $\bf D_F$ both register zero photon. It is
interesting to note that the failure of the standard measurement
to distinguish the four quasi-Bell states, $\left(\left| \alpha
\right\rangle \left| -\alpha \right\rangle \mp \left| -\alpha
\right\rangle \left| \alpha \right\rangle \right)$ and
$\left(\left| \alpha \right\rangle \left|\alpha \right\rangle \mp
\left| -\alpha \right\rangle \left| -\alpha \right\rangle
\right)$, in case of maximal entanglement is also subject to the
same condition, i.e., the two detectors $\bf D_E$ and $\bf D_F$
registering no photon\cite{11,12}. Nevertheless, the success
probability decreases as the degree of entanglement decreases,
because the probability for the two detectors $\bf D_E$ and $\bf
D_F$ to detect no photon increases as the parameter $\beta \left(
= \alpha\tan\eta \right)$ decreases.
\begin{figure}[t]
\includegraphics[width=7cm]{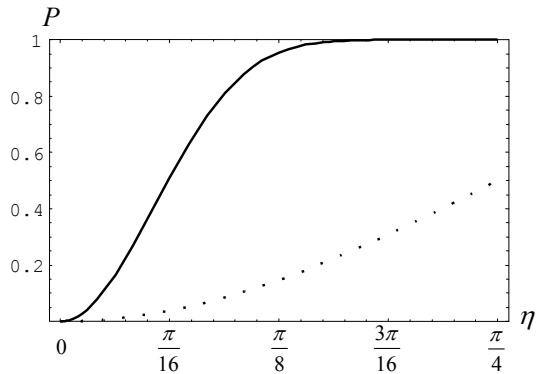}% Here is how to import EPS art
\caption{\label{f6} The success probability $P$ vs.~$\eta$ for
case (c) with $\alpha=3$ (solid curve) and for cases (a) and (b)
(dotted curve). }
\end{figure}
We show in Fig.\ref{f6} the success probability as a function of
$\eta$ for case (c) where $\alpha=3$ is assumed, along with the
curve $\sin^2 \eta$ representing the success probability for cases
(a) and (b). One first notes that the success probability for case
(c) is nearly 1 when $\eta = {\pi \over 4}$, whereas that for
cases (a) and (b) is at most 1/2, as long as only linear optical
schemes are employed. Furthermore, the success probability for
case (c) decreases much more slowly than that for cases (a) and
(b), as $\eta$ is decreased from $\pi \over 4$. In other words,
the success probability for case (c) remains much closer to its
maximum value for the same amount of amplitude errors. This
robustness of the success probability with respect to amplitude
errors stems largely from the robustness of the degree of
entanglement with respect to amplitude errors. In fig.\ref{f7} we
show the degree of entanglement, $E=-\mbox{Tr}\left( \rho \log
\rho \right)$, with respect to $\eta$ for the nonmaximally
entangled states (\ref{e7}) and (\ref{e1}). The degree of
entanglement remains close to 1 for a relatively broad range of
$\eta$ for state (\ref{e7}), while it decrease rapidly as $\eta$
is decreased from $\pi \over 4$ for state (\ref{e1}). Fig.\ref{f8}
shows the success probability as a function of the degree of
entanglement for case (c) and for cases (a) and (b). Regardless of
the value of the degree of entanglement, the success probability
for case (c) is roughly twice that for cases (a) and (b). We
mention that the behavior of the success probability for case (c)
exhibited in Figs.\ref{f6} and \ref{f8} remains essentially the
same, as long as $\alpha \gtrsim 3$. We also mention that the
success probability for case (c) varies little with respect to the
unknown amplitudes $x$ and
$y$, as long as $\alpha \gtrsim 3$.\\
\indent It should be noted that our observation that the success
probability for case (c) remains close to 1 over a relatively
broad range of the amount of amplitude error stands only if we
have photon detectors which can distinguish between even and odd
numbers of photons. At present, photon detectors with single
photon resolution which can distinguish between $n$ and $\left(
n+1 \right)$ photons do not exist, despite recent efforts and
progress\cite{15,16}. It is certainly a high technical challenge
to construct such detectors. It is, however, in principle possible
to distinguish between $n$ and $(n+1)$ photons, if one uses photon
counters that have been developed to distinguish between no photon
and a single photon or between a single photon and two photons
with high quantum efficiency\cite{15} in an arrangement of
detector cascades or $N$-ports\cite{17}. It has also been
suggested that $n$ and $(n+1)$ photons can be distinguished by
utilizing homodyne detection looking at the imaginary
quadrature\cite{18} or by coupling the field to a two-level atom
through nonlinear interaction\cite{12,19}.

In conclusion, we have presented experimental schemes to perform
generalized measurements for conclusive teleportation when the
sender and the receiver share nonmaximally entangled states for
three different cases. Our analysis shows that the scheme with a
coherent state qubit is relatively robust against amplitude
errors. This advantage may play an important role in actual
implementation of linear optical schemes for quantum communication
and quantum computation\cite {18,20,21} based on coherent-state
qubits.

\begin{acknowledgments}
This research was supported by Korea Research Foundation under
Contract No. 2002-070-C00029.
\end{acknowledgments}

\end{document}